# Shock Drift Acceleration of Ions in an Interplanetary Shock Observed by MMS

E.L.M. Hanson[1,2], O.V. Agapitov[2], I.Y. Vasko[2,3], F.S. Mozer[2], V. Krasnoselskikh[2,4], S.D. Bale[2], L. Avanov[5,6], Y. Khotyaintsev[7], B. Giles[6]

## ABSTRACT

An interplanetary (IP) shock wave was recorded crossing the Magnetospheric Multiscale (MMS) constellation on 2018 January 8. Plasma measurements upstream of the shock indicate efficient proton acceleration in the IP shock ramp: 2-7 keV protons are observed upstream for about three minutes (~8000 km) ahead of the IP shock ramp, outrunning the upstream waves. The differential energy flux (DEF) of 2-7 keV protons decays slowly with distance from the ramp towards the upstream region (dropping by about half within 8 Earth radii from the ramp) and is lessened by a factor of about four in the downstream compared to the ramp (within a distance comparable to the gyroradius of ~keV protons). Comparison with test-particle simulations has confirmed that the mechanism accelerating the solar wind protons and injecting them upstream is classical shock drift acceleration. This example of observed proton acceleration by a low-Mach, quasi-perpendicular shock may be applicable to astrophysical contexts, such as supernova remnants or the acceleration of cosmic rays.

[1] Corresponding author lily.hanson@berkeley.edu
[2] Space Sciences Laboratory, University of California, Berkeley, CA 94720
[3] Space Research Institute of Russian Academy of Sciences, Moscow, Russia
[4] LPC2E/CNRS-University of Orléans, Orléans, France
[5] Astronomy Department, University of Maryland, College Park, MD, USA
[6] NASA Goddard Space Flight Center, Greenbelt, MD, USA
[7] Swedish Institute of Space Physics, Uppsala, Sweden

# 1. INTRODUCTION

Collisionless shocks, which effect the transition from upstream to downstream plasma states over a length scale far smaller than the collisional mean free path, appear throughout the universe and play an important role in shaping the emissions of supernovae and accelerating cosmic rays (Biermann et al. 2018; Gargaté & Spitkovsky 2012). They also feature prominently in the heliosphere, for instance as planetary bow shocks and interplanetary (IP) shocks. Their prevalence has historically made planetary environments and the solar wind valuable hunting grounds for shock studies that are based upon *in situ* measurements, from space missions such as ISEE (Sckopke et al. 1983), INTERSHOCK (Walker et al. 1999), AMPTE (Balikhin et al. 1999), Cluster (Bale et al. 2005; Krasnoselskikh et al. 2013; Kruparova et al. 2019), THEMIS (Wilson et al. 2014a, 2014b), Polar (Bale & Mozer 2007; Hull et al. 2006, 2012), and Wind (Wilson et al. 2012, 2017).

One of the primary consequences of passage through a shock transition is deceleration and increased thermalization of the incoming ions. The cross-shock electric field potential impacts the distribution functions of both ions and electrons (Goodrich & Scudder 1984) and may contribute to ion reflection, but hitherto investigations of the electric field and electrostatic potential in collisionless shocks have been sparse due to the challenges of making *in situ* measurements. Fortunately, recent missions such as Magnetospheric Multiscale (MMS) are expanding our horizons with observations of unprecedented resolution (Burch et al. 2016; Fuselier et al. 2016).

Gradients within the shock transition may give rise to the phenomenon of shock drift acceleration (SDA), where ions travel along the shock front and gyrate through the ramp multiple times before either passing into the downstream or escaping into the upstream region (Armstrong et al. 1985; Decker & Vlahos 1985; Pesses et al. 1982). When ions have been accelerated by SDA, they may pass downstream or escape upstream, but the highest-energy ions are expected to be found upstream (Armstrong et al. 1985). SDA is distinct from but has been included in models for diffusive shock acceleration (DSA) because both mechanisms act on particles drifting along the shock front within a small distance (on the order of the ion gyroradius) of the ramp (Jokipii 1982). Ion acceleration models that rely upon time spent within the strong field gradients of the shock transition are not limited to the solar wind but have also been discussed within the context of interstellar pick-up ions and the heliospheric termination shock (Lee et al. 1996; Zank et al. 1996). Both SDA and DSA are invoked in more general astrophysical cases as well, particularly to explain observations of cosmic rays and supernova remnants (Biermann et al. 2018; Caprioli & Spitkovsky 2014; Gargaté & Spitkovsky 2012; Ohira 2016; Park et al. 2015; Zank et al. 2015). Lario et al. (2019) have described IP shock observations where enhancements in ions less than 10 keV appeared close to the shock. Similarly, ion populations of ~20 keV associated with an IP shock have been recorded by the ARTEMIS spacecraft (Kajdič et al. 2017). In this paper we report experimental evidence of SDA in a population of 2-7 keV ions observed ahead of an interplanetary (IP) shock.

## 2. DATA

The MMS mission boasts a full complement of high-resolution instruments for *in situ* plasma measurements (Burch et al. 2016; Fuselier et al. 2016). Ion and electron distributions and moments were obtained from the Fast Plasma Investigation (FPI) (Pollock et al. 2016). We checked the heavy ion observations from the Hot Plasma Composition Analyzer (HPCA) (Young et al. 2016), but within our interval of interest there was insufficient data for us to draw conclusions. The Fluxgate Magnetometer (FGM) provided magnetic field measurements (Russell et al. 2016; Torbert et al. 2016), while high-frequency (AC), 3D electric fields are measured by the Axial and Spin-Plane Double Probe electric field instruments (Ergun et al. 2016; Lindqvist et al. 2016).

## 3. OVERVIEW OF THE INTERPLANETARY SHOCK

The four MMS spacecraft recorded an interplanetary (IP) shock on 2018 January 8 (magnetic field data and solar wind bulk velocity are shown in Figure 1a and 1b, respectively). The IP shock crossed the spacecraft at 330 km s$^{-1}$ in the spacecraft frame at 06:41:11. The distance between individual MMS probes ranged from 15 km to 24 km at the time of observation (Hanson et al. 2019). An overview of the Fast Survey and Burst data is shown in Figure 1. The cross-shock potential for this quasi-perpendicular, low-Mach shock has been estimated to be near 25V based on electric field measurements (Cohen et al. 2019; Hanson et al. 2019). This IP shock was recorded later by the two ARTEMIS spacecraft (Angelopoulos 2011; Auster et al. 2008), and the combined observations of MMS and ARTEMIS suggested that the shock front was nearly planar on a spatial scale of 60 Earth radii ($R_E$) (Hanson et al. 2019). The angle between the shock normal and the upstream magnetic field was 69°, and the magnetosonic Mach number was 1.1 (1.5 if ion temperatures from Wind are used in the calculation) (Hanson et al. 2019).

A population of the upstream ions reflected by the shock (less than 5% of the total upstream solar wind density) consisted of nearly-specularly reflected ions (Cohen et al. 2019) and ions accelerated to energies between 3 keV and 7 keV (Hanson et al. 2019). The specularly reflected ions, which appear for a few tenths of a second in the upstream region ~150 km ahead of the shock ramp (highlighted by the frame in Figure 1h), are seen as an enhancement in the Sun-directed flux of ions in the energy flux near 1 keV, described in detail by Cohen et al. (2019).

The ion population of interest to this paper (highlighted in Figure 1e) is seen as a broad, faint band between 2 keV and 5-7 keV, slowly decaying with distance from the ramp in the upstream direction until it reaches the noise level. It is strongest during 06:40:00-06:41:15 but is still clearly present as early as 06:37:50, several RE ahead of the shock. The upper bound of energy increases closer to the shock front. The flux of ions in the accelerated energy range are decreases by a factor of about four downstream, within a distance comparable to the Larmor radius of a 2-7 keV proton. Here, a brief enhancement of 0.2-2 keV ion flux is observed (Figures 1e, f, h). Observations of the accelerated ions in the upstream so far from the shock were possible due to the planar shock structure, narrow ramp, and relatively small angle between the upstream magnetic field and the shock normal (Cohen et al. 2019; Hanson et al. 2019), as well as the steady upstream magnetic field. Enhancements of a similar energy range but lower flux are visible even farther ahead of the ramp in Figure 1b (e.g. 06:27-06:33), but it is unclear whether these are due to the shock or to other solar wind phenomena.

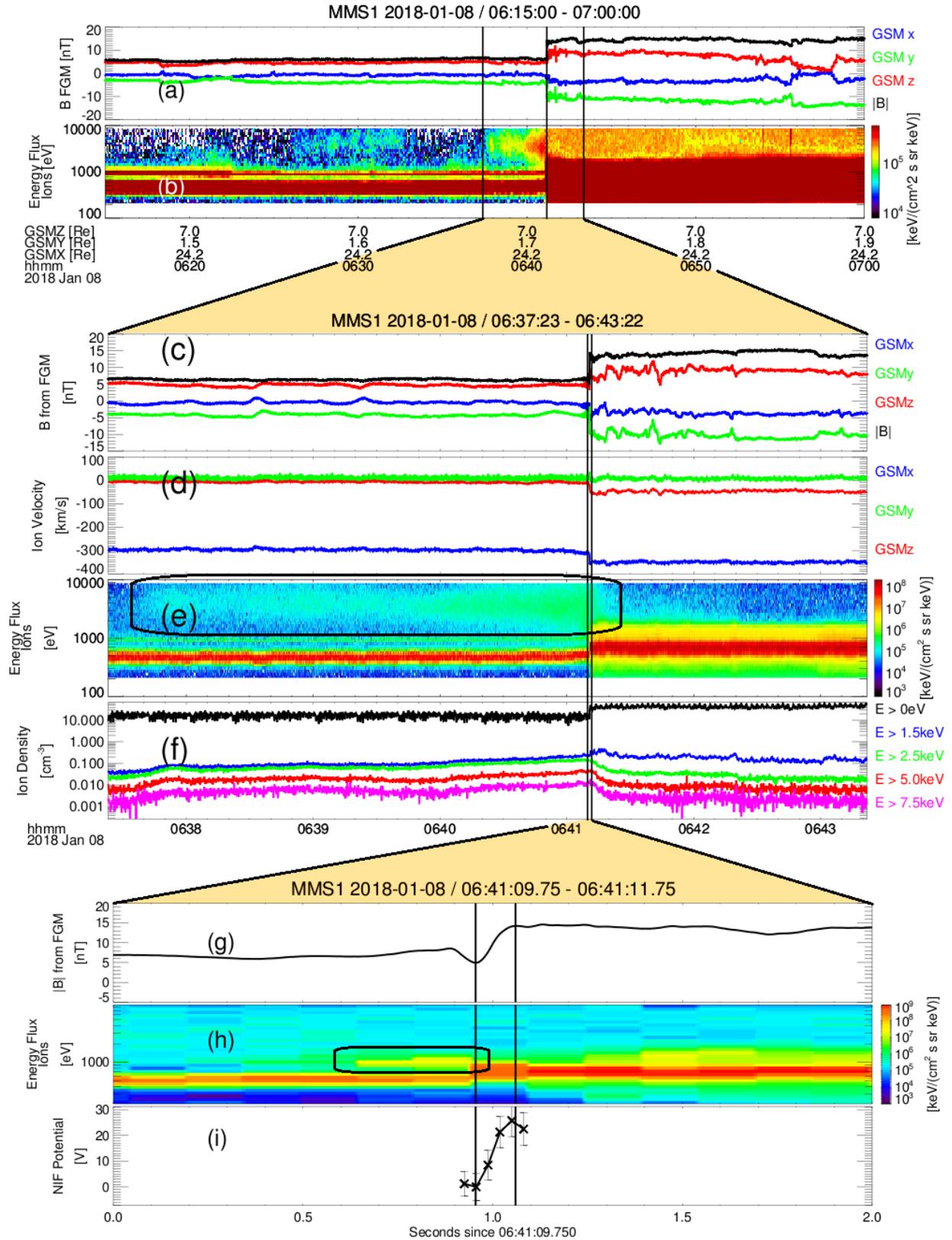

**Figure 1.** Summary of IP shock observation by MMS1 in Geocentric Solar Magnetic (GSM)

coordinates. Panels (a)-(b) show Fast Survey data in the time range 06:15-07:00, panels (c)-(f) show the burst-rate data, which covers 06:37:23-06:43:22, and panels (g)-(i) present 2s of burst-rate data around the ramp. (a) and (c) Magnetic field vector and magnitude; (b), (e), (h) ion energy flux with different color scales; (d) ion velocity vector; (f) ion density split by several minimum energy thresholds; (g) magnetic field magnitude; (i) cross-shock potential.

Figure 2 illustrates the wave fields in the vicinity of the shock front for the magnetic field magnitude and the high-frequency electric field component parallel to the magnetic field. In Figure 2b, which shows the wavelet of the magnetic field GSM x-component, low-frequency whistler waves (0.5-2Hz) are evident, extending for about 2000 km on either side of the shock front. The strongest fluctuations are centered near 1Hz, and the phase (group) speed is 184 km s$^{-1}$ (370 km s$^{-1}$). It is important to note that the group speed is comparable to the shock speed of 330 km s$^{-1}$ because this enables the whistler waves to continue running ahead of the ramp without escaping easily.

Both wavelets indicate significant activity near the shock: large-scale whistler wave precursors (electromagnetic) just upstream and electrostatic wave activity in the ramp. The electrostatic waves most likely represent some combination of ion-acoustic waves, lower hybrid waves, and bipolar structures with wavelengths on the order of a few Debye lengths (Gurnett 1985; Hull et al. 2006; Vasko et al. 2018). Although the electrostatic waves may contribute to the pitch angle scattering of electrons (Vasko et al. 2018), they are highly inefficient for scattering or acceleration protons. In some cases, whistler waves may lead to efficient acceleration of protons (Kis et al. 2013). For this particular shock, protons cannot be accelerated by the observed whistler and electrostatic waves. The magnetic field and plasma pressure gradients are insufficient for phase trapping in the whistler wave (Artemyev et al. 2013), while the wavelength of the electrostatic waves (~1 km) differs by about two orders of magnitude from the proton Larmor radius (~300 km), rendering trapping and acceleration impossible.

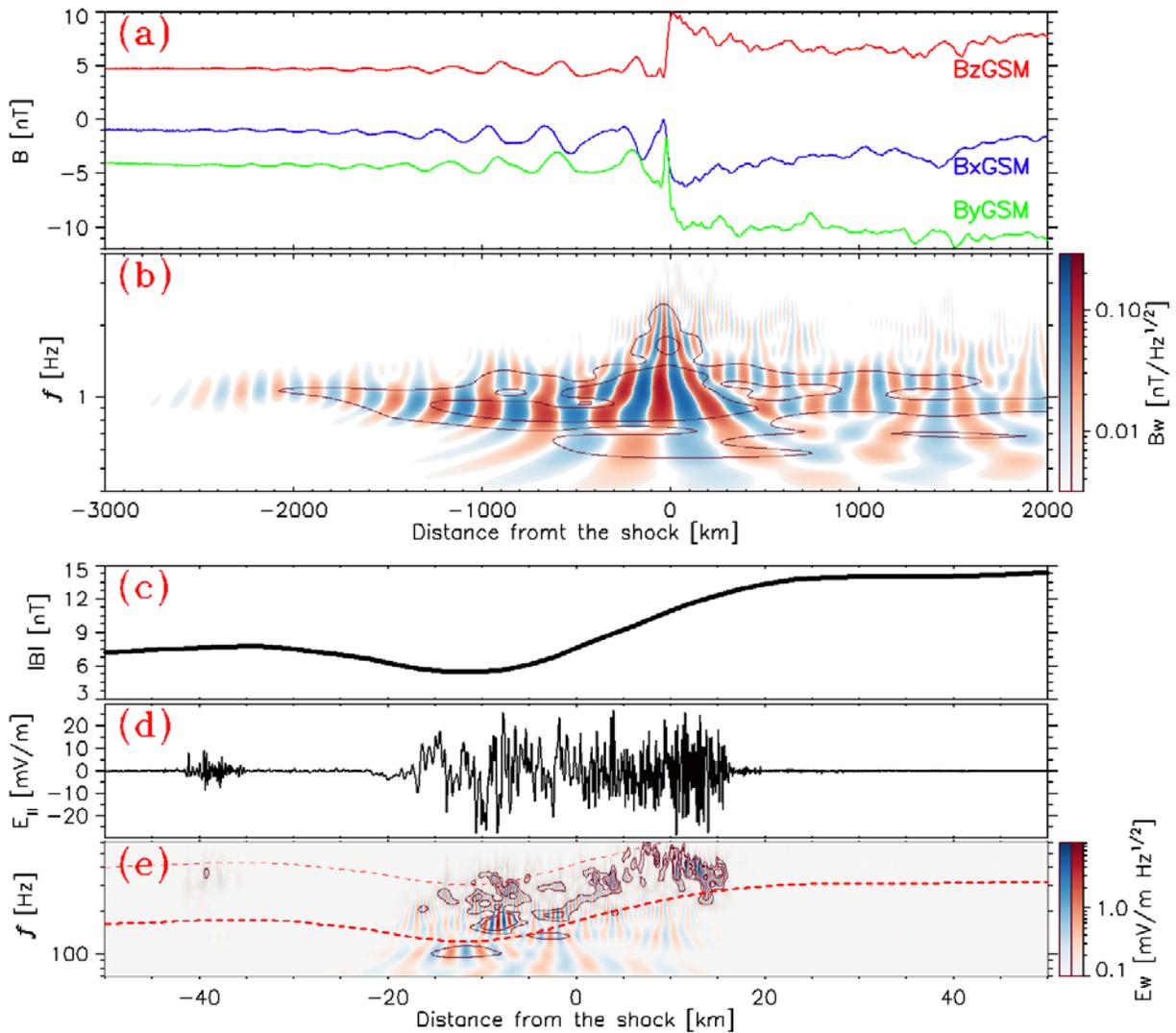

**Figure 2.** Wave activity surrounding the shock front in the spacecraft frame from MMS4 observations. Time has been scaled by the shock speed to present distance from the shock in the horizontal axes. Panels (a) and (b) extend 3000 km ahead of the shock and 2000 km behind it, and Panels (c)-(e) span 50 km on either side of the shock. (a) GSM components of magnetic field; (b) Morlet wavelet of the GSM x component of the magnetic field, showing whistler waves surrounding the shock. The wavelet amplitude is represented by the black contours, while red (blue) indicate positive (negative) values of the wavelet's real part; (c) magnetic field magnitude; (d) component of AC electric field parallel to the magnetic field; (e) Morlet wavelet of the parallel electric field, indicating lower hybrid wave activity within the ramp. The contours and color scale

are the same as described in Panel (b). The red dashed (dotted) curve indicates 0.8 (2) times the electron cyclotron frequency.

## 4. ACCELERATED IONS PRECEDING THE IP SHOCK

The ion density, subdivided by minimum energy thresholds in Figure 1f, shows that ions with energy in excess of 1.5 keV comprise less than 1-2% of the total ion density throughout the observation. While the total ion density upstream of the shock remains nearly constant at 20 cm$^{-3}$, the higher-energy ions experience increasing density starting from about 06:37:50 (3min 20s ahead of the ramp). For all but the highest-energy ions (those with energy greater than 7.5 keV), this increase is monotonic until the shock front is encountered. The density of the highest-energy ions dips temporarily around 06:39:20-06:40:20 (1min 50s to 50s ahead of the ramp).

The density of all ions with energy exceeding 2.5 keV drops rapidly (on the scale of the ion gyroradius) once the shock front has passed. The density profile for ions with energy greater than 1.5 keV behaves differently: for ten seconds after the passage of the shock, the density jumps by a factor of two. Ions in the energy range 1.5-2.5 keV, presumably a combination of accelerated protons and compressed alpha particles, must be responsible for this phenomenon since no such jump is evident in the profiles for energies over 2.5 keV.

The panels of Figure 3 show three different plots for each of three time periods surrounding the shock. Figures 3a, 3d, and 3g present data from a far upstream region, before the accelerated ions appear (3min 47s to 3min 25s ahead of the ramp, or 06:37:23-45UT). Figures 3b, 3e, and 3h show observations taken in the near upstream region when the accelerated ions are most strongly visible (1min 10s to 9s ahead of the ramp, or 06:40:00-41:01UT). Finally, the downstream region is shown in Figures 3c, 3f, and 3i (49s to 2min 11s behind the ramp, or 06:42:00-43:22UT). Figure 3a-c presents profiles of ion differential energy flux (DEF) versus energy for three distinct time periods surrounding the shock. Figure 3d-f shows the angular distribution of ions in the Despun spacecraft Body Coordinate System (DBCS; (Pollock et al. 2016)), where particles appearing in 0° or 360° in azimuth have come from the direction of the Sun, and the ecliptic plane is near 90° in polar angle. The filled contours indicate where ions of a given energy flux exceed a minimum threshold. In Figure 3g-i, the ion DEF is shown in pitch angle versus energy.

The far upstream observations are typical of quiet solar wind conditions. In Figure 3a, the incoming solar wind population is dominated by a narrow beam of protons in the energy range 400-500 eV, while the alpha particles form a peak two orders of magnitude smaller in the energy range 0.9-1 keV. In Figure 3d, only the narrow, localized solar wind beam is evident, coming towards the spacecraft from the Sun. Similarly, in Figure 3g the solar wind protons form a strong, concentrated beam near a pitch angle of 90° in all panels, with a weaker beam at the same angle but higher energy for the alpha particles.

The downstream panels hold no surprises either. After the passage of the shock front, the DEF profile has changed significantly, as is shown in Figure 3c. Here the dominant proton beam has increased in peak DEF and broadened in energy to 600-700 eV. The alpha particles no longer form a separate beam but are subsumed in a higher-energy shoulder jutting out from the broadened solar wind beam. The DEF drops off rapidly with increasing energy beyond about 1.5 keV. Figure 3f shows that the solar wind beam is still the dominant population, but it has a much larger angular spread than in Figure 3d. This broadening is also evident in Figure 3i, where the solar wind beam peaks at the same pitch angles as in Figure 3g but is significantly spread in both pitch angle and energy, with a higher peak energy.

It is in the upstream panels nearest the ramp that the most interesting features appear. In Figure 3b, the solar wind proton and alpha particle peaks are nearly the same as in Figure 3a, but a broad beam has appeared in energies above 1 keV. For particles with energies between 2 keV and 5 keV, the total density is near 0.5 cm$^{-3}$. The DEF of this near upstream, high-energy peak is less than the solar wind protons by two orders of magnitude but comparable to the alpha particles, and greater than the far-upstream ions of the same energy by a factor of 3-5. Figure 3e shows that the solar wind beam has been joined by a new population of high-energy ions, moving in an azimuthal direction similar to the solar wind but offset in polar angle. The high-energy ions also exhibit a larger angular spread in comparison with the incoming solar wind. The solar wind and helium populations of Figure 3h look much the same as in Figure 3g, but in Figure 3h, the accelerated ions are seen as an enhancement for pitch angles between 90° and 150° and energies above ~2 keV. Note that the accelerated ions are more persistent at angles more aligned with the magnetic field, and only the highest-energy ions are able to access more perpendicular pitch angles. Particles that undergo shock drift acceleration escape along the direction of the magnetic field before they are returned to the solar wind population.

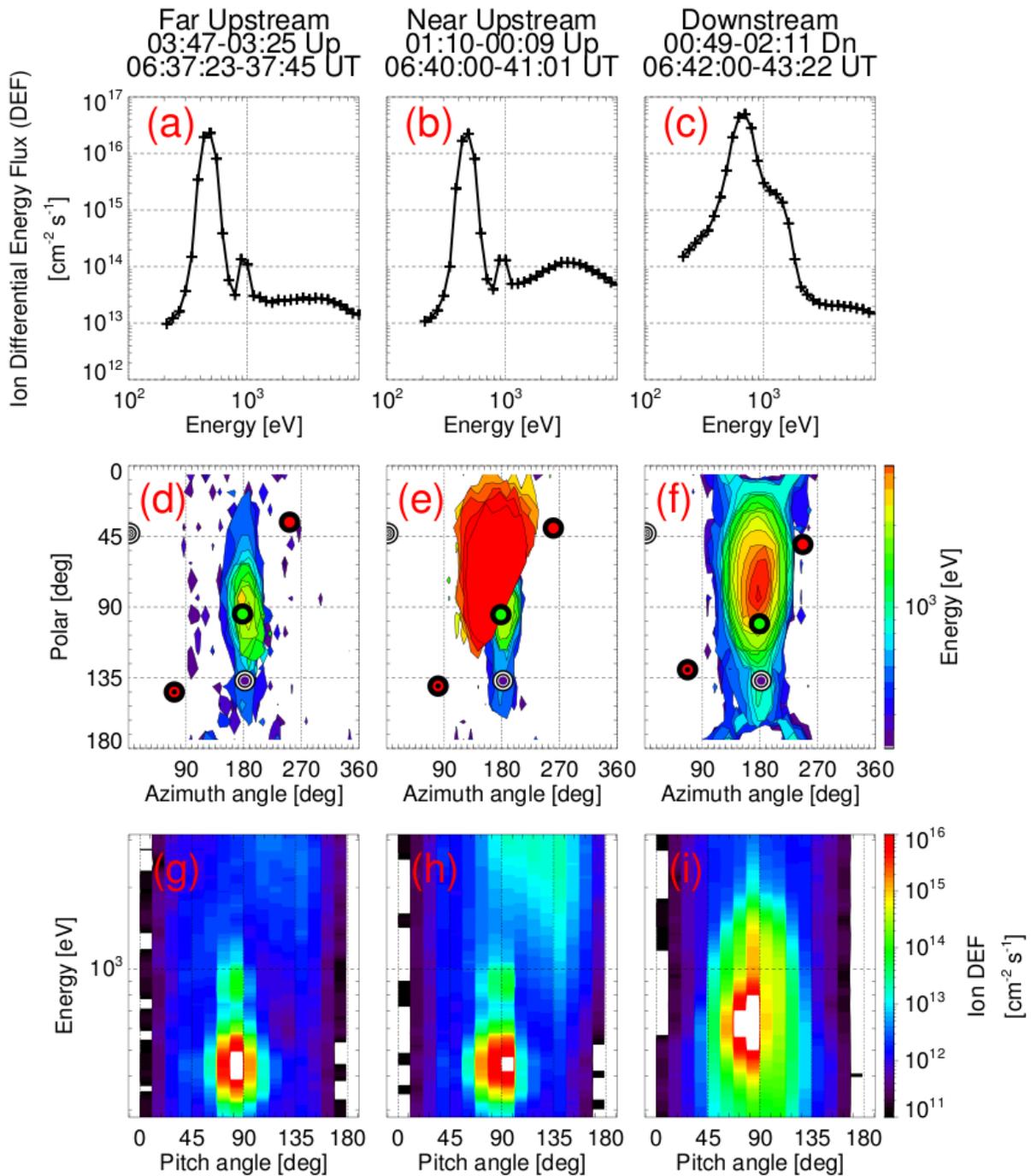

**Figure 3.** Panels (a)-(c) are profiles of DEF versus energy. Panels (d)-(f) show the angular locations where the energy flux at a given energy exceeds a minimum threshold. The coordinates are azimuthal angle (x-axis) and polar angle (y-axis) in the Despun spacecraft Body Coordinate System (DBCS). Panels (g)-(i) show ion distributions in energy and pitch angle with respect to the magnetic field. Panels (a), (d), and (g) display data averaged over a far upstream time range (06:37:23-45 UT, or 3min 47s to 3min 25s ahead of the shock); panels (b), (e), and (h) show the near upstream (06:40:00-41:01 UT, or 1min 10s to 0min 9s ahead of the shock); and panels (c), (f), and (i) show the downstream (06:42:00-43:22 UT, or 0min 49s to 2min 11s after the shock). In panels (d)-(f), the azimuthal angle 180° points away from the Sun, and the polar angle 90° is approximately aligned with the ecliptic plane. The red circle indicates the direction of the magnetic field, while the red circle with a small black dot in the center shows the opposite. Similarly, the green dot at the right edge of the plot represents the average solar wind direction. The shock normal is represented by a purple dot surrounded by concentric black and white rings, and black and white rings with no central dot indicate the opposite direction.

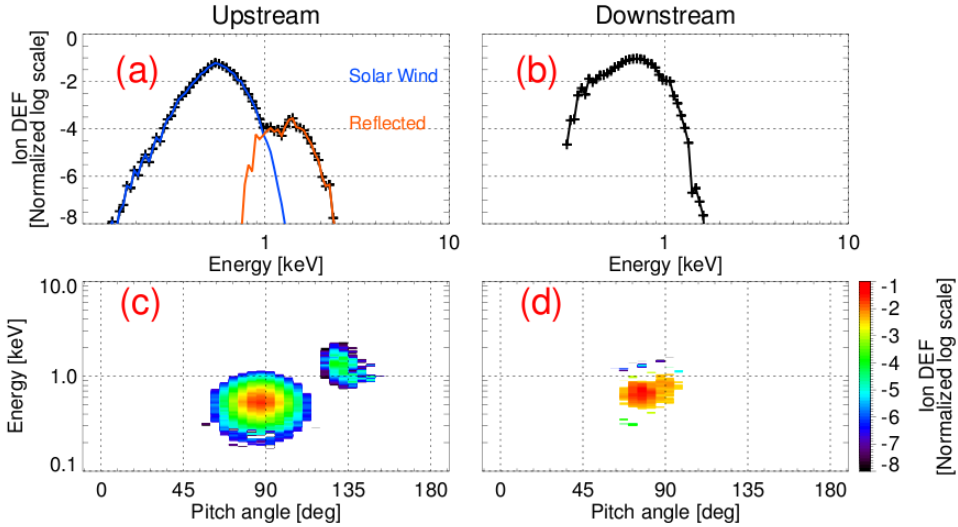

**Figure 4.** Panels (a)-(b) show the proton DEF profiles in the simulation upstream and downstream regions, respectively; compare the upstream (downstream) panel to Figure 3b (3c). The blue curve is the incoming solar wind, and the orange curve is the reflected population, which peaks at 1-2 keV. Panels (c)-(d) give the proton DEF versus energy and pitch angle for the upstream and downstream regions, respectively; compare the upstream (downstream) panel to Figure 3h (3i). The reflected protons are concentrated around pitch angles of 120°-135°. Simulation DEF units are logarithmic and normalized to the total DEF of the downstream protons.

We performed a test particle simulation with the observed shock parameters to verify that SDA can explain the accelerated protons' energy and pitch angle distributions. From the simulation data, we obtained the phase space density and DEF of protons in the spacecraft frame (see the Appendix for a detailed description). The DEF of incoming protons and protons accelerated by SDA and escaping upstream is shown in Figure 4a, while Figure 4b presents the DEF of protons in the downstream region. The distributions in Figures 4a-b are in a similar format as the experimental data in Figures 3a-c. We can see that the protons accelerated by the SDA are observed as a peak of DEF at energies of 1-2 keV and their DEF is 2-3 orders of magnitude less than the incoming population, consistent with the observations in Figure 3b. We stress that the actual acceleration of protons is up to 7 keV (see Figure A1b in the Appendix), but for the considered Maxwellian distribution of the incoming protons, the phase space density and DEF of these protons is quite small. Figures 4c-d present the DEF of protons in the upstream and downstream regions in dependence on the energy and pitch angle. Comparison to Figures 3h-i shows that the simulations well reproduce the DEF of incoming protons and protons in the downstream regions, i.e. the peak of the DEF is around 90° and around 500 eV in the upstream region and around 700 eV in the downstream region. Most interestingly, the simulations show that the protons accelerated by SDA and escaping to the upstream region are observed in a narrow range of pitch angles of 120°-135°, which is quite consistent with the observations in Figure 3h. The test-particle simulations demonstrate that the upstream protons are accelerated within the shock front by the SDA process explaining both the energies of the accelerated protons as well as their narrow pitch angle distribution.

## 5. SUMMARY AND CONCLUSIONS

An interplanetary (IP) plasma shock wave was recorded by the four MMS spacecraft on 2018 January 8. The plasma measurements provide evidence of ion acceleration from ~0.5 keV (solar wind ions) to 2-7 keV. These accelerated ions are observed to be injected into the plasma upstream of the IP shock, preceding the ramp by about three minutes (~8000 km) and decaying slowly in the upstream direction. In the downstream region, the flux of 2-7 keV ions falls by a factor of about four compared to the flux just before the ramp. Their energy range exceeds that of the solar wind protons and alpha particles, and though they travel in nearly the same azimuthal direction with respect to the spacecraft as the solar wind beam, they are slightly offset in polar angle and exhibit a broader angular distribution.

The shock itself is surrounded by wave activity: in addition to whistler precursors, there are short-scale electrostatic waves in the upstream and downstream regions as well as in the ramp. These waves were processed from the perspective of their impact on ions, and it was determined that the waves cannot provide the observed acceleration. The electrostatic waves have wavelengths too small compared to the ion gyroradius to provide the observed acceleration, and acceleration due to whistlers would require a large population of reflected particles at low energies.

The mechanism of shock drift acceleration, in which the ions are incrementally accelerated along the component of the electric field parallel to the shock front during repeated crossings of the ramp, has been shown to be capable of accelerating the 2-7 keV ions and reflecting them upstream. The pitch angle distribution of the accelerated ion population (the higher-energy ions have larger pitch angles) is consistent with this mechanism; similar features are also seen in the test-particle simulation.

We note that ions are accelerated by a factor of five to ten in their interaction with this low-Mach, quasi-perpendicular IP shock. Since Mach numbers for astrophysical shocks are generally much higher, correspondingly greater acceleration of ions should be possible.

## Acknowledgments

The authors thank the entire MMS team for providing such excellent data, which is publicly available at: https://lasp.colorado.edu/mms/sdc/public/about/how-to/. Work at UC Berkeley benefited from the support of NASA contract NNN06AA01C and NSF grant number 1914670. The work of I.V. was supported by NASA MMS Guest Investigator grant No. 80NSSC18K0155. I.V. thanks for support the International Space Science Institute (ISSI), Switzerland, Bern. The authors would like to thank the anonymous referee for insightful comments.

Appendix A. TEST-PARTICLE SIMULATIONS

To address the origin of a few keV protons observed in a relatively narrow range of pitch angles upstream of the shock, we have performed a test-particle simulation of the interaction of incoming protons with the shock magnetic field and cross-shock electrostatic field. The shock is assumed to be planar with the x-axis along the shock normal directed into the upstream region, the z-axis in the plane of the shock and along the major magnetic field component $B_z$, and the y-axis completing the right-hand coordinate system. The magnetic field of the shock is modelled with two components, $\boldsymbol{B} = (B_x, 0, B_z)$, where $B_x = B_u \sin\theta_{Bn}$ =const and $B_z(x) = B_u \cos\theta_{Bn} + 0.5\Delta B(1 - \tanh(x/h_B))$, where $B_u \sim 6$nT is the magnetic field magnitude in the upstream region ($x \to +\infty$), $\theta_{Bn} \sim 115°$ is the angle between the x-axis and the magnetic field in the upstream region, $\Delta B \sim 9$nT is the magnetic field jump across the shock, and $h_B$ is the gradient scale of the magnetic field. The electrostatic cross-shock field is modelled with $E_x = E_0 \cosh^{-2}(x/h_E)$, where $E_0 \sim 1.1$mV/m is the peak value of the cross-shock electrostatic field and $h_E$ is the gradient scale of the cross-shock electrostatic field. The gradients scales $h_B = 8$km and $h_E = 14$km were estimated by fitting the measured magnetic field and electrostatic field profiles to the model profiles.

The simulation was performed in the Normal Incidence Frame (NIF). In that frame protons are incoming into the shock with the bulk velocity of about Vn~90 km/s in the direction opposite to the normal (i.e. toward –x direction) and the corresponding motional electric field of 0.5 mV/m is directed in the –y direction. To model the distribution function of incoming protons we run 200 thousand protons from a far upstream region and follow proton trajectories until they get either into a far downstream region or back into the far upstream region. The distribution function of the incoming protons is assumed to be an isotropic Maxwellian drifting with velocity 90 km/s opposite to the normal and temperature of 10 eV, which is rather close to the proton temperature estimate by the Wind spacecraft located in the solar wind about 200 $R_E$ upstream of MMS. The results of the simulations are transformed into the spacecraft frame to facilitate the comparison with the MMS measurements.

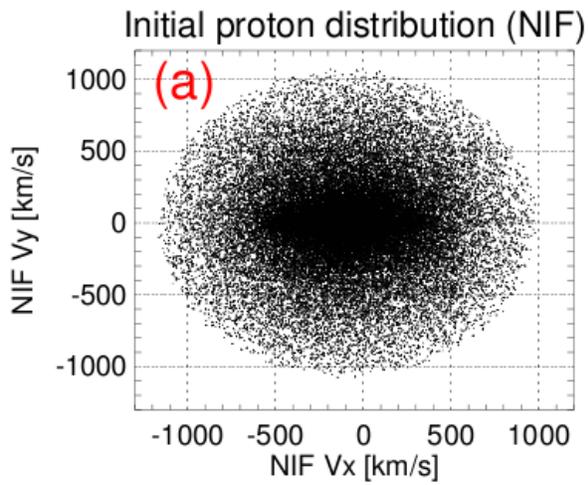
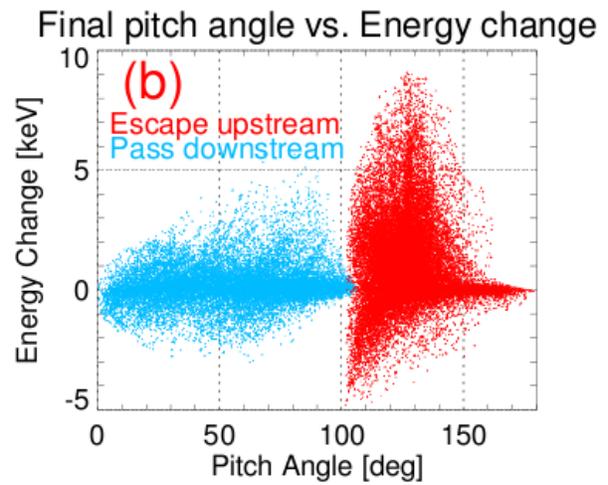
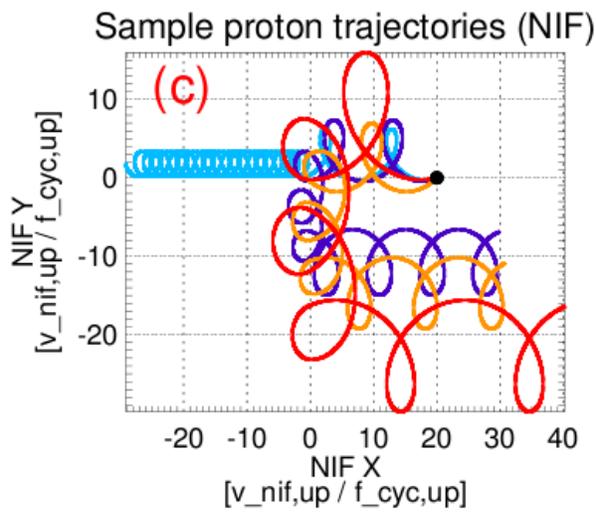
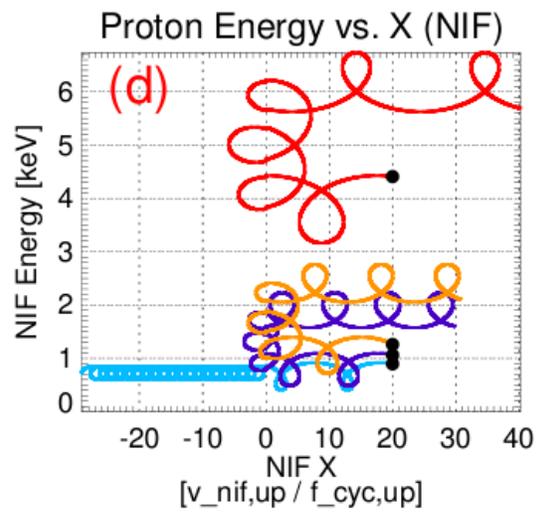
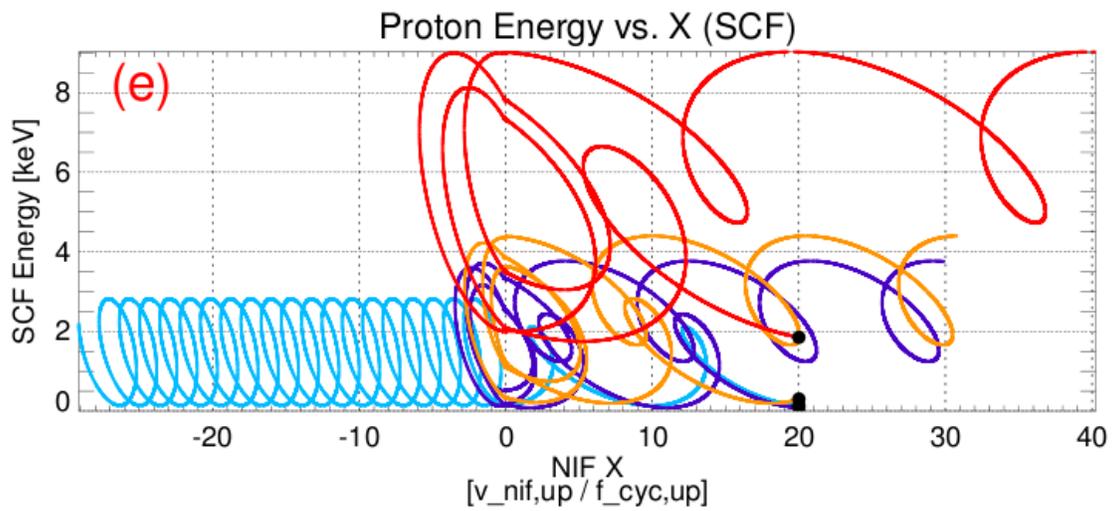

**Figure A1.** Summary of test-particle simulation results. All panels but (e) show data in NIF. Panel (a): The distribution of velocities $V_x$ and $V_y$ of incoming protons, corresponding to 200 thousand protons with velocities distributed uniformly in a sphere centered at $V_x$~-90km/s. Panel (b): Change in energy (keV) versus final pitch angle (degrees). Protons that escape to the upstream are indicated with red dots, while protons transmitted downstream are light blue. Panel (c): Trajectories of four sample protons in the NIF x-y plane. The protons are launched from x=20, while the shock ramp is at x=0. Distances are scaled by the ratio of the average upstream NIF velocity and the corresponding proton cyclotron frequency. Panel (d): Proton energy in the NIF (keV) versus position x for the same four protons as in panel (c). Panel (e): Proton energy in the spacecraft-frame (keV) versus position x for the same four protons as in panels (c) and (d).

Figure A1 demonstrates the results of the test-particle simulation. Panel (a) presents the distribution of initial velocities of launched protons in the ($V_x$,$V_y$)-plane. Panel (b) presents the distribution of the energy change and pitch angles of the protons after interaction with the shock as viewed in the spacecraft frame. Protons ending up in the far downstream are observed at all pitch angles from 0° to ~90° and accelerated by up to a few keV. Protons escaping back to the far upstream region are observed in a narrow range of pitch angles, 120°-150°, and can be seen to be accelerated by up to 7 keV. Panel (c) presents trajectories of several protons interacting with the shock; the blue trace indicates a proton that is transmitted to the downstream, while violet, orange, and red traces represent protons that finally escape to the far upstream region. The same color scheme is employed in panels (d) and (e). Panel (d) shows the evolution of proton energy as viewed in the NIF, while the energy evolution in the spacecraft frame is shown in panel (e); both demonstrate that the acceleration experienced by the protons is due to classical shock drift acceleration (SDA), i.e. protons drift along the front of the shock and are accelerated by the motional electric field of 0.5 mV/m.

The trajectories of the test protons allow us to compute the phase space density of protons escaping back to the far upstream and ending up in the downstream regions of the shock. For that purpose we use the Liouville's theorem and consider each of 200 thousand protons as a macro-particle carrying the phase space density determined by the initial phase space density of incoming protons. The phase space density is then used to compute the DEF of each proton population.